\documentclass[12pt,preprint]{aastex}

\shorttitle{Radiation Hydrodynamics in SPH}
\shortauthors{Viau et al.}

\begin{document}

\title{An implicit method for radiative transfer
      with the diffusion approximation in SPH}

\author{Serge Viau, Pierre Bastien}
\affil{D\'epartement de Physique and Observatoire du Mont--M\'egantic,
Universit\'e de Montr\'eal, C.P. 6128, Succ. centre--ville,
Montr\'eal, Quebec, Canada}
\email{serge.viau@collanaud.qc.ca; bastien@astro.umontreal.ca}

\and

\author{Seung-Hoon Cha}
\affil{D\'epartement de Physique and Observatoire du Mont--M\'egantic,
Universit\'e de Montr\'eal, C.P. 6128, Succ. centre--ville,
Montr\'eal, Quebec, Canada,\\and\\
Korea Astronomy and Space Science Institute, 61-1, Whaam-Dong, Youseong-Gu, Taejeon, Korea}
\email{cha@kasi.re.kr}

\begin{abstract}
An implicit method for radiative transfer in SPH is described.
The diffusion approximation is used, and
the hydrodynamic calculations are performed by a fully
three--dimensional SPH code. Instead of the energy equation of
state for an ideal gas, 
various energy states and the dissociation of hydrogen molecules
are considered in the energy calculation for a more realistic
temperature and pressure determination. In order to test the implicit
code, we have performed non--isothermal collapse simulations of a
centrally condensed cloud, and have compared our results with 
those of finite difference calculations performed by 
MB93. The results produced by the two completely different
numerical methods agree well with each other.
\end{abstract}


\section{Introduction}
In order to understand the various stages 
of star formation
it is essential to follow the exact thermal evolution of a
collapsing cloud because some important dynamical processes,
for example fragmentation \citep{0521,0524,0519,0518,0637,0661},
are closely related to the thermal
evolution. The thermal evolution of a collapsing cloud is
described by radiation hydrodynamics, and inevitably involves
a numerical solution. However, it is not easy to implement and
run a multi--dimensional radiation hydrodynamic code, so various
approximate
methods have been used instead of solving the full
radiation hydrodynamic equations directly.

\citet{0528,0544,0263,0269} used one--dimensional
hydrodynamic codes including a more realistic approach to
radiative transfer. Such an approach can track the exact
thermal evolution of a collapsing
cloud, and is easily compared to the observational results.
However, in a one--dimensional code it is impossible to
observe some important dynamics, for example fragmentation.

On the other hand, a simplified treatment for the radiative transfer is needed
in multi--dimensional hydrodynamic codes.
The Eddington approximation was used by \citet{0533,0534,0514,0292},
while \citet{0265} used the diffusion approximation in his
one--dimensional simulations.

However, the multi-dimensional calculations with radiative transfer
published so far have been done mostly with codes using the finite
difference method.
Although smooth particle hydrodynamics
(hereafter SPH) codes are now used quite commonly in the
calculations for self--gravitating clouds, so far
few attempts
\footnote{\cite{0662} was published after the original work
of \citet{t0011},
so we need to emphasize the differences between
their approach and ours. First of all, \citet{0662} concentrated on
one--dimensional tests while this paper presents a fully three--dimensional
calculation. Our test calculations are shown in Section 3.
Furthermore, a more realistic energy calculation has been used in our
code rather than the ideal equation of state. Details about the energy
calculation are presented in Section 2. On the other hand, \citet{0662}
used a more detailed treatment of the radiative transfer which allows
the radiation and gas temperature to be different.}
to include radiative transfer in an SPH code have been published
\citep{0170,0515,0662}.
The main reason is that the diffusion
approximation for treating radiative transfer there has a double
differential in spatial coordinates which is very difficult to
evaluate accurately with SPH because the particles occupy
in principle arbitrary positions. \citet[][hereafter B85]{0515}
derived an equation
which bypasses the double differential by converting it to a
single differential. \citet[][hereafter CM99]{0660} extended
the B85 method to treat cases with a discontinuous conductivity.

Another difficulty in the treatment of radiative transfer is the
large difference in time scales. The radiative time scale is
much shorter than the dynamical time scale in a collapsing
cloud. Therefore, an implicit numerical scheme is needed to
treat the thermal and dynamical evolutions simultaneously.
\citet{t0011} developed an effective implicit scheme for
radiative transfer in a fully three--dimensional SPH code
after many unsuccessful attempts.

We have performed a non--isothermal cloud collapse
with the implicit code, and compared our results
with those of \citet[][hereafter MB93]{0292}. The comparison should be useful
because the two methods are completely different from each other.

The implicit scheme which combines the diffusion approximation
and SPH is described in section 2 in detail.
We have used a more realistic energy calculation instead of
an ideal equation of state for the determination of temperature,
and the detailed procedure for the specific internal energy
calculation is also given in the same section.
The non--isothermal cloud collapse simulations
and the comparison of our results with those of
MB93 are given in section 3. The summary is in section 4.

\section{Numerical methods}
\subsection{Smoothed Particle Hydrodynamics : SPH}
SPH \citep{0170,0019} is a grid--free and fully
Lagrangian method, and therefore has been
widely used in gravitational collapse simulations. The Lagrangian
hydrodynamics with self--gravitation are given by 
\begin{eqnarray}
\label{lhd1}
\frac{D\rho}{Dt} &=& -\rho\nabla\cdot\mathbf v,\\ 
\label{lhd2}
\frac{D\mathbf v}{Dt} &=& -\frac{1}{\rho}
\mathbf\nabla P-\mathbf\nabla^2\Phi,\\ 
\label{lhd3}
\frac{Du}{Dt} &=& -\frac{P}{\rho}\nabla\cdot\mathbf v,
\end{eqnarray}
where $D/Dt$ is the Lagrangian derivative, $u$ is the specific
internal energy and other variables have their usual meaning.
Although there are many alternatives for the SPH formulation of
equations (\ref{lhd1}) -- (\ref{lhd3}) \citep{0229,0222},
we have used a common form given by
\begin{eqnarray}
\label{sph1}
\rho &=& \sum_j m_j W_{ij},\\
\label{sph2}
\frac{\Delta\mathbf v}{\Delta t} &=& -\sum_jm_j
\left(\frac{P_i}{\rho^2_i}+\frac{P_j}{\rho^2_j}+\Pi_{ij}\right)
\mathbf\nabla_iW_{ij} -\mathbf\nabla^2\Phi_i,\\
\label{sph3}
\frac{\Delta u}{\Delta t} &=& \frac{1}{2}\sum_jm_j
\left(\frac{P_i}{\rho^2_i}+\frac{P_j}{\rho^2_j}+\Pi_{ij}\right)
(\mathbf v_i-\mathbf v_j)\cdot \mathbf\nabla_iW_{ij}.
\end{eqnarray}
Here $\Pi_{ij}$ is the artificial viscosity given by 
\begin{equation}
\label{avterm}
\frac{-\alpha \bar c_{s,ij}\mu_{ij}+\beta\mu^2_{ij}}{\bar\rho_{ij}},
\end{equation}
where $\mu_{ij}=h\frac{\mathbf v_{ij}\cdot\mathbf
r_{ij}}{r^2_{ij}+\eta^2}$, $h$ is the smoothing length,
$\mathbf v_{ij}=\mathbf v_i-\mathbf
v_j$, $\mathbf r_{ij}=\mathbf r_i-\mathbf r_j$,
$\bar c_{s,ij}=(c_{s,i}+c_{s,j})/2$, $\bar\rho_{ij}=(\rho_i+\rho_j)/2$, and
$\alpha$, $\beta$ and $\eta$ are free parameters\footnote{We have used
$\alpha=1.0$, $\beta=2.0$ and $\eta=0.1$ in all simulations.
$\eta$ is not a critical free parameter \citep{0123} for the artificial
viscosity, because normally $\mathbf r_i\ne\mathbf r_j$.}.
Here $c_s$ is the
sound speed. Although the artificial viscosity has a
side--effect in differentially rotating systems \citep{0223,0160,0500},
it is essential for the treatment of shock waves in SPH.
$W$ in equations (\ref{sph1}) - (\ref{sph3}) is a
kernel function, and the M4 kernel \citep{0007} has been used in
our simulations. For additional details about SPH, the reader
is referred to the reviews by \citet{b0003} and \citet{0229}.

\subsection{Diffusion approximation in SPH}
The diffusion approximation has been adopted in our
three-dimensional SPH code for
the treatment of radiative transfer, because
it is very hard to trace the exact behavior of individual photons
in the multi--dimensional hydrodynamic code \citep[but see][]{0642}.
The diffusion approximation is strictly valid only for regions
of great optical depth (i.e. optically thick medium).
However, the early stage of the collapse of a cloud is optically
thin because of the very low density and effective cooling.
\citet{0265} discussed this point in his one-dimensional
simulations, and concluded that the diffusion approximation
can be applied in the isothermal stage of a collapsing cloud,
because the diffusion approximation essentially keeps
the cloud temperature at the boundary value in the early stage
of the collapse.

The energy equation with the diffusion term is given by
\begin{equation}
\label{ene}
\frac{Du}{Dt} = -\frac{P}{\rho} \nabla \cdot \mathbf{v}
- \frac{1}{\rho}\nabla\cdot\mathbf F, 
\end{equation}
where $\mathbf F$ is the radiative flux, and is given by
\begin{equation}
\label{radflux}
\mathbf F = -\frac{4acT^3}{3\kappa_R\rho}\nabla T,
\end{equation}
where $\kappa_R$ is the Rosseland mean opacity,
$a$ is $\frac{4\sigma_{SB}}{c}$, $c$ is the speed of light and
$\sigma_{SB}$ is the Stephan--Boltzmann constant.
If an effective conductivity,
$Q(\equiv -\frac{16\sigma_{SB} T^3}{3\kappa_R\rho})$ is defined,
equation (\ref{ene}) may be rewritten as
\begin{equation}
\label{newene}
\frac{Du}{Dt} = -\frac{P}{\rho} \nabla \cdot \mathbf{v} -
\frac{1}{\rho}\nabla\cdot (Q\nabla T).
\end{equation}
Equation (\ref{newene}) contains a double derivative, and this
$\mathbf\nabla^2$
operation is very sensitive to the disorder in the particle
distribution, such that it may cause numerical noise in the simulation.
B85 suggested another formulation
for the diffusion term to avoid this $\mathbf\nabla^2$ operator.
We describe here briefly the treatment of B85.

In one--dimension, the diffusion term of equation (\ref{newene}) becomes
\begin{equation}
\label{onedimdif}
\frac{1}{\rho}\left(\frac{dQ}{dx}\frac{dT}{dx} +
Q\frac{d^2T}{dx^2}\right).
\end{equation}
The derivation for the first term of
equation (\ref{onedimdif}) starts from the Taylor expansions of
$T(x')$ and $Q(x')$ around $T(x)$ and $Q(x)$, respectively,
\begin{eqnarray}
\label{taylorq}
Q(x') &=& Q(x) + (x'-x)\frac{dQ}{dx} +
\frac{1}{2}(x'-x)^2\frac{d^2Q}{dx^2} + \cdots,\\
\label{taylort}
T(x') &=& T(x) + (x'-x)\frac{dT}{dx} +
\frac{1}{2}(x'-x)^2\frac{d^2T}{dx^2} + \cdots.
\end{eqnarray}
From equations (\ref{taylorq}) and (\ref{taylort}) one can
derive,
\begin{equation}
\label{middqdt}
[Q(x')-Q(x)][T(x')-T(x)] = (x'-x)^2\frac{dQ}{dx}\frac{dT}{dx},
\end{equation}
and the integral interpolant form of equation (\ref{middqdt}) becomes
\begin{equation}
\label{firstterm}
\frac{dQ}{dx}\frac{dT}{dx} = 
-\int\frac{[Q(x')-Q(x)][T(x')-T(x)]}{x-x'}
\frac{\partial W(x-x')}{\partial x}dx'.
\end{equation}

For the derivation of the second term of equation (\ref{onedimdif}),
we will use equation (\ref{taylort}),
\begin{equation}
-\int\frac{T(x')-T(x)}{x-x'}\frac{\partial W(x-x')}{\partial x}dx'
= \int\frac{dT}{dx}\frac{\partial W(x-x')}{\partial x}dx' 
-\frac{1}{2}\int(x-x')\frac{d^2T}{dx^2}
\frac{\partial W(x-x')}{\partial x}dx',
\end{equation}
and it becomes 
\begin{equation}
\label{secondterm}
Q\frac{d^2T}{dx^2} =
-2Q\int\frac{T(x')-T(x)}{x-x'}\frac{\partial W(x-x')}{\partial x}dx'.
\end{equation}
From equations (\ref{firstterm}) and (\ref{secondterm}),
the diffusion term of equation (\ref{ene}) in one--dimension becomes
\begin{equation}
\label{diffsph1}
\frac{1}{\rho}\left(\frac{dQ}{dx}\frac{dT}{dx}+Q\frac{d^2T}{dx^2}\right)
= \sum_j\frac{m_j}{\rho_i\rho_j}\frac{(Q_i+Q_j)(T_i-T_j)}{x_i-x_j}
\frac{\partial W_{ij}}{\partial x_i},
\end{equation}
and the three-dimensional form becomes
\begin{equation}
\label{diffsph3}
\frac{1}{\rho}
\left(\mathbf\nabla Q\cdot \mathbf\nabla T + Q\mathbf\nabla^2T\right)
= \sum_j\frac{m_j}{\rho_i\rho_j}
\frac{(Q_i+Q_j)(T_i-T_j)}{|\mathbf r_i-\mathbf r_j|^2}
(\mathbf r_i-\mathbf r_j)\cdot\mathbf\nabla_iW_{ij}.
\end{equation}
With equation (\ref{diffsph1}) or (\ref{diffsph3}),
the calculation of $\mathbf\nabla^2$
can be avoided, and the energy equation becomes finally
\begin{equation}
\label{finalene}
\frac{\Delta u}{\Delta t} = \frac{1}{2}\sum_jm_j
\left(\frac{P_i}{\rho^2_i}+\frac{P_j}{\rho^2_j}+\Pi_{ij}\right)
(\mathbf v_i-\mathbf v_j)\cdot \mathbf\nabla_iW_{ij}
+\sum_j\frac{m_j}{\rho_i\rho_j}
\frac{(Q_i+Q_j)(T_i-T_j)}{|\mathbf r_i-\mathbf r_j|^2}
(\mathbf r_i-\mathbf r_j)\cdot\mathbf\nabla_iW_{ij}.
\end{equation}

This method derived by B85 is valid for a constant or smoothly varing
conductivity. CM99 suggested a modification to treat media with a
discontinuous conductivity.
They considered the continuity of conductive heat flux across the border of
adjacent cells (or particles), and found a better expression, 
$\frac{4Q_iQ_j}{Q_i+Q_j}$, for the effective
conductivity term instead of $(Q_i+Q_j)$ in equation (\ref{finalene}). 
CM99 also performed various tests with the new formulation, and some of
them are repeated in the next section.

\subsection{Test of the new formulation: thermal conduction}
\label{heatsec}
Given the approximation involved in the derivation
presented above, it is not obvious that a solution of 
equation (\ref{finalene}) is also a solution of
equation (\ref{ene}). Therefore
we have performed a test of the thermal conduction
problem to verify the treatment for the double
derivative.
The thermal diffusion equation is given by 
\begin{equation}
\label{heatdiffusion}
\rho\frac{du}{dt} = \nabla\cdot(\kappa\nabla T),
\end{equation}
where $t$ is time and $\kappa$ is the thermal conductivity.
The simplest case for the thermal diffusion equation is that $\kappa$ is
a constant (i.e. a homogeneous medium). In this case
equation (\ref{heatdiffusion}) becomes
\begin{equation}
\label{hdsimple}
\frac{dT}{dt} = \frac{\kappa}{\rho\sigma}\nabla^2T.
\end{equation}
Here a simple relation between $u$ and $T$, $u=\sigma T$ is assumed, and 
$\sigma$ is the specific heat of the medium. The treatments of B85 and CM99
are the same in this simple case, and so equation (\ref{hdsimple})
can be rewritten as
\begin{equation}
\label{heatnumsol}
\frac{\Delta T}{\Delta t} =
\frac{2\kappa}{\rho_i\sigma}\sum_j\frac{m_j}{\rho_j}\frac{T_i-T_j}{x_i-x_j}
\frac{\partial W_{ij}}{\partial x_i}
\end{equation}
in one--dimension.

\begin{figure}[htbp]
\centering
\includegraphics[scale=0.4]{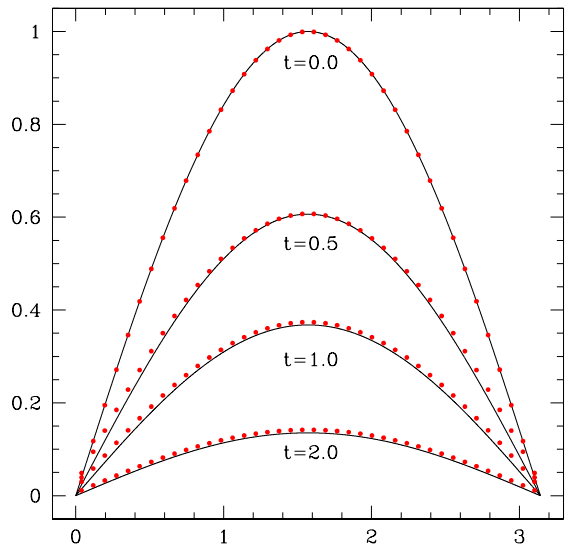}
\caption{A numerical solution for the heat conduction equation at $t =
0.0, 1.0, 1.5$ and $2.0$. The dots are the numerical solution
using equation (\ref{heatnumsol}), and
the solid lines are the analytic solution given by
equation (\ref{heatsol}).
There is very good agreement between these two solutions.
The boundary temperature is set to $0$K in this simulation.
\label{cont}}
\end{figure}

\begin{figure}[htbp]
\centering
\includegraphics[scale=0.4]{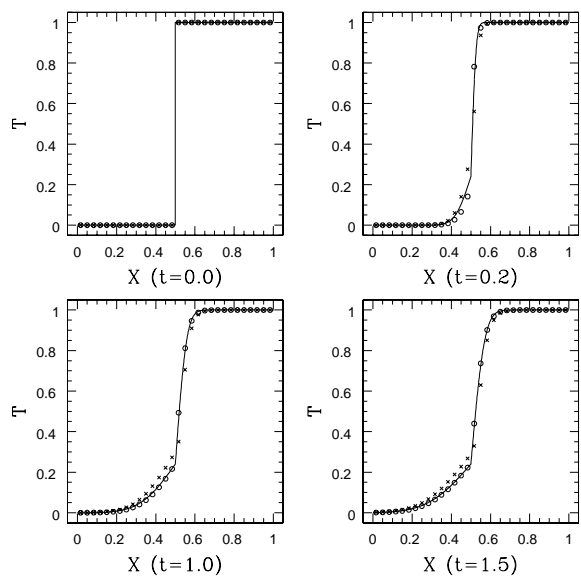}
\caption{Numerical solutions for the heat conduction equation 
with different conductivity at
$t = 0.0, 0.2, 1.0$ and $1.5$. The open circles are 
with the expression of CM99, and
the crosses are with the original method of B85, respectively.
The analytic solution is drawn by a solid line. 
From $t=0.2$, the numerical
results show small deviations from the analytic solution.
Generally speaking, the modification of CM99 shows
a better agreement than B85.
\label{cmresult}}
\end{figure}

For this test,
we have used 40 particles in $0\le x\le \pi$, and a constant
$h$ value for all particles. $\rho$, $\sigma$ and $\kappa$ are set to unity,
and all particles keep their original positions in this test, so the
situation is very similar to a one--dimensional finite
difference simulation. The temperature distribution at $t=0$
is given by
\begin{equation}
\label{heatinit}
T(x,0) = \sin x,
\end{equation}
and the boundary temperature is set to $0$K.
The analytic solution of equation (\ref{hdsimple}) is,
\begin{equation}
\label{heatsol}
T(x,t) = e^{-t}\sin x.
\end{equation}
Figure \ref{cont} shows the results until $t=2$.
The solid lines in the figure are the analytic solution
given by equation (\ref{heatsol}) at different times,
and the dots are the results of our numerical
calculation. The numerical solution reproduces the analytic
solution very well and confirms the validity of the
B85 formulation.
Furthermore we have performed several tests for cases with a 
discontinuous conductivity as
presented in CM99. We will show one of them. In this test the medium
is divided into two parts, and each part has different values for the
conductivity, so the
resultant diffusivity $\left(\equiv\frac{\kappa}{\rho\sigma}\right)$ is
different in the left and right part of the medium.
The initial conditions for this test are shown in Table \ref{cmtest}.
Refer to section 5.3 of CM99 for details of this test.
\begin{table}
\caption{Initial conditions for discontinuous conductivity test}
\label{cmtest}
\begin{center}
\begin{tabular}{ccc}
\hline
 & left & right \\
\hline
$T$ & 0 & 1 \\
$\rho$ & 1 & 1 \\
$\kappa$ & 10 & 1 \\
$\sigma$ & 1 & 1 \\
\hline
\end{tabular}
\end{center}
\medskip
\end{table}
Figure \ref{cmresult} shows the results. The method of B85 as modified
by CM99 (open circles) is more accurate than the original one
by B85 (crosses). We have used the the modification suggested by CM99 
in our three--dimensional test (See section 3).
\subsection{Implicit scheme for radiative transfer}
\label{implicitsec}
The radiative time scale is much shorter
than the dynamical time scale in a collapsing cloud,
so the numerical integration of equation (\ref{finalene})
is not easy. An implicit scheme has
been developed to treat the radiative transfer and the dynamical
evolution simultaneously \citep{t0011}, and we explain
each step of this implicit scheme.

\begin{enumerate}
\renewcommand{\theenumi}{\arabic{enumi}}
\item Define a function ${\mathcal F}$  from equation (\ref{ene}),
\begin{equation}
\label{f1}
{\mathcal F} = u^*_i - u^0_i +
\delta t\left(\frac{P}{\rho} \nabla \cdot \mathbf{v}\right)_i^0 +
\delta t\left(\frac{1}{\rho}\nabla\cdot\mathbf F\right)_i^*,
\end{equation}
where $u^0_i$ is the former step value, and $u^*_i$ is the
new value updated by the iteration.
The second and third terms of equation (\ref{f1}) are known for each time
step. The goal of the iteration is to find the value of $u_i^*$ (or
equivalently $T_i^*$) which will bring ${\mathcal F}=0$.
\item Set the boundary values for temperature, $T_L$ and $T_R$.
These boundary temperatures are initially limited by the
temperature range of the opacity table (see section
\ref{rosselandsec}).
\item Find $u_L$ and $u_R$ from $T_L$ and $T_R$, respectively. At this
stage various energy states and the dissociation of 
hydrogen molecules are considered (see section
\ref{thermosec}).
\item Find the median value, $T_i^*$ using the bisection method
or the Van Wijingaarden--Dekker--Brent method \citep[e.g.][]{b0012}.
The convergence speed of the Van Wijingaarden--Dekker--Brent method
is higher than that of the bisection method.
Typically $\sim 10$ iterations
are required to solve equation (\ref{f1}) for one particle. This
iteration is started using $u_i^* = u_i^0$ as the initial guess. 
Rapid convergence here is a concern since we are in a double loop in the code.
\item Find $u_i^*$ from $T_i^*$
\item After all values of $u_i^*$ have been found, we compute ${\mathcal F}$
again using equation (\ref{f1}) with the new $u_i^*$.
Note that only the first and fourth terms change in equantion (\ref{f1}).
\item Repeat steps 4--6 until ${\mathcal F} < tol$ for all particles,
using the latest $u_i^*$ found for each particle.
\end{enumerate}
The tolerance, $tol$ in step 6 is closely related to the resolution
(and speed as well) of the
calculation, and set to $10^{-5}$ in our simulations.
Typically several iterations $(\le 10)$ are required to reach the desired
precision.

In the procedure explained above, note that
only the temperature and specific
internal energy are updated at every iteration, while the pressure
and density are fixed. The Rosseland mean opacity, $\kappa_R$ is
also updated by the corresponding temperature and fixed density.
This whole procedure is executed for each particle separately,
i.e. the new temperature is determined for each particle while
keeping all the other particles at their original temperature.
The convergence
according to the same criterion has to be satisfied for all
particles. If not, the above procedure is repeated until it is
satisfied.

For comparison, the classical method to solve for the gravitational force
for the N--body problem involves $N^2/2$ steps. Using a tree
approach brings this to $N\log N$. Here the radiative flux equation has
to be solved simultaneously for all particles. When the specific energy
for one particle is varied, all the other ones are fixed. In summary,
this method is in $N^2$ steps, times the number of iterations required.
This is the price to pay for the different time scales between the radiative
and dynamical events. But the method converges well and yields very good
results. The number of iterations required allows taking larger time steps.

As an example, we have used a small Altix (Intel Itanium2 processor) for the 
calculations with 50000 particles reported below. The results were obtained 
with our serial code version, and the calculation time is just about an hour. 
We stopped the calculation when the maximum density (this happens usually at 
the cloud center)
reaches $10^5$ times the initial density.
Generally speaking, the RHD calculation takes $\approx$10 times more CPU time
than the HD calculation without radiative transfer.

The detailed procedures to derive $\kappa_R$ and the specific
internal energy from a given temperature will be described in
sections \ref{rosselandsec} and \ref{thermosec}, respectively.

\subsection{Rosseland mean opacity}
\label{rosselandsec}
\begin{figure}[htbp]
\centering
\includegraphics[scale=0.4]{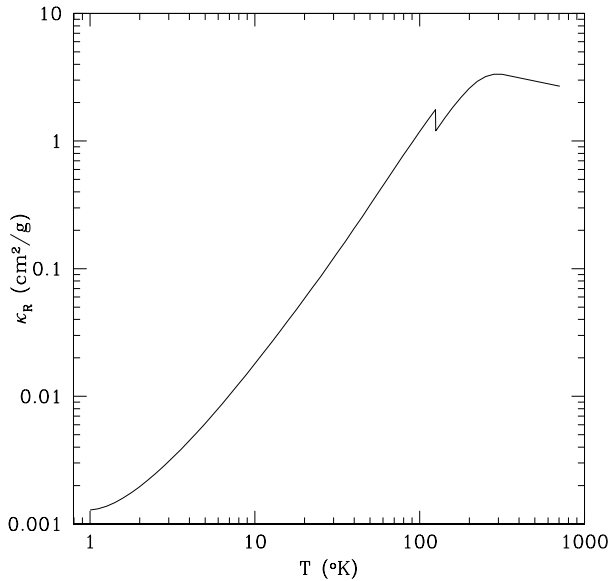}
\caption{The Rosseland mean opacity for $1$K $<T<708 $K. For 
$T<316$K, we have used the model of \citet{0528} with the recent
values for the extinction factor of \citet{0529}. For 
$316$K $<T<708$K, we have used simple interpolated values. The
jump around $T \simeq 125$K is due to the sublimation of the ice
mantles.\label{kr1}}
\end{figure}

\begin{figure}[htbp]
\centering
\includegraphics[scale=0.4]{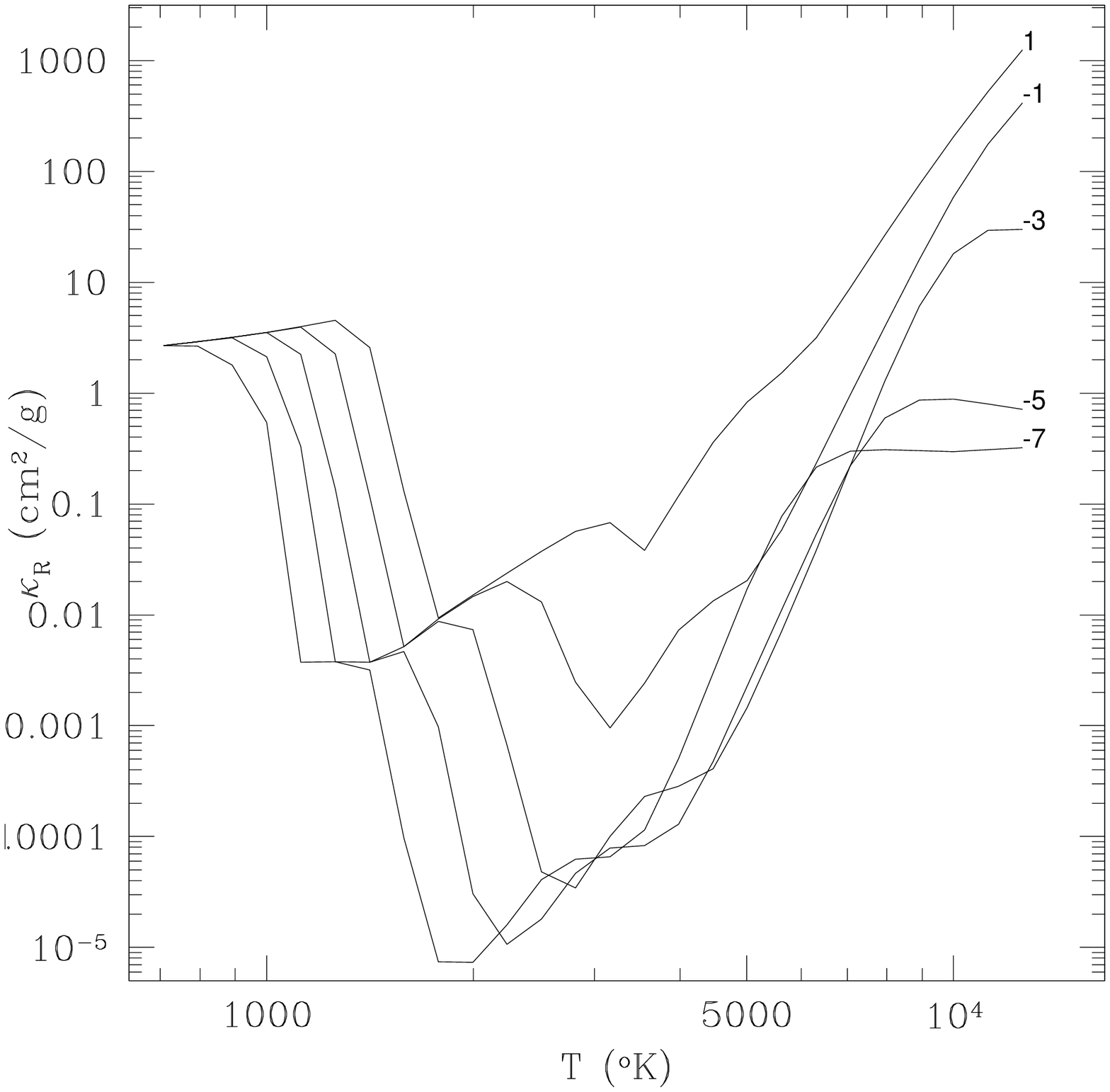}
\caption{The Rosseland mean opacity as a function of temperature
for $708$K $<T<12500$K and $-7\le\log R\le1$ \citep{0530}.
Each curve is labeled
with the value of $\log R$. There are jumps at
$T\simeq 1200$K and $2000$K. They are due to the sublimation of
silicates and amorphous carbon, respectively. All dust components
evaporate at $T\simeq 2000$K, and the absorption of molecular and
atomic lines become the dominant coolant.\label{kr2}}
\end{figure}

The Rosseland mean opacity, $\kappa_R$ should be determined in
order to implement radiative transfer in the diffusion approximation.
The definition of the Rosseland mean opacity is given by
\begin{equation}
\label{kappar}
\frac{1}{\kappa_R} =
\frac{\int^\infty_0\frac{1}{\kappa_\nu}\frac{dB_\nu}{dT}d\nu}
{\frac{dB}{dT}},
\end{equation}
where $B_\nu$ is the blackbody function at a frequency $\nu$
and $B$ is the frequency integrated blackbody function
$(\equiv\int^\infty_0B_\nu d\nu)$. $\kappa_\nu$ in equation
(\ref{kappar}) is determined by
\begin{equation}
\label{kappanu}
\kappa_\nu = \frac{1}{\rho}nQ_{ext}\pi r_d^2,
\end{equation}
where $n$ is the number density, $Q_{ext}$ and $r_d$ are the extinction
factor and size of dust grains.
In the lower temperature range $(T<316$K$)$, 
we have used the model for $\kappa_\nu$ developed by \citet{0528} with 
the more recent values for the extinction factor,
$Q_{ext}$ provided by \citet{0529}.
The Rosseland mean opacity for this lower temperature region is
shown in Figure \ref{kr1}. In the figure, the opacity
jumps at $T\simeq 125$K because of the sublimation of the ice mantle
on dust grains.
In the higher temperature region $(708$K $< T < 12500$K$)$,
we have used 
the model of \citet{0530} for $\kappa_R$,
which considered the absorption of
atomic lines (with more than 8 million lines) and molecular lines
(with nearly 60 million lines). Grain absorption and scattering
due to silicates, iron, carbon and SiC have also been considered 
in this model. \citet{0530} tabulated the opacity using a parameter,
$R$, defined by
\begin{equation}
\label{logr}
R = \frac{\rho}{T^3_6},
\end{equation}
where $T^3_6$ is $T/10^6$.
Figure \ref{kr2} shows $\kappa_R$ in the ranges $-7\le\log R\le1$ and
$708$K $< T < 12500$K.
There are two jumps at $T\simeq 1200$K and $2000$K
due to the sublimation of silicates and amorphous carbon, respectively.
Therefore, all dust components evaporate at $T\simeq 2000$K, and
the absorption of molecular and atomic lines becomes dominant
above that temperature.
Each curve in the figure is labeled with the value of $\log R$.
For the intermediate temperature region $(316$K $<T<708$K$)$,
we have used
simple interpolated values for $\kappa_R$ (see Figure \ref{kr1}).

\subsection{Thermodynamics}
\label{thermosec}
Temperature has been used as an independent variable in the
iteration explained in section \ref{implicitsec}, and
the specific internal energy of each particle is
evaluated at every step of the iteration. The specific internal
energy can be determined uniquely for a given temperature, density
and chemical composition. For a more exact determination,
the specific internal energies of hydrogen, helium
and metals are calculated separately. We also consider
the dissociation of hydrogen molecules in the energy calculation,
but the ionization of hydrogen atoms is omitted, because
ionization is negligible in the temperature range corresponding to our
simulations. We now explain how to derive the
energy. In this description, $X$, $Y$ and $Z$
denote the mass fractions of hydrogen, helium and metals, respectively. 

The total specific internal energy is given by
\begin{equation}
\label{utotal}
u = u(H) + u(H_2) + u(H_{2diss}) + u(He) + u(M).
\end{equation}
Here $u(H)$ is the specific internal energy of hydrogen atoms, 
given by
\begin{equation}
u(H) = \frac{3}{2}Xy\frac{kT}{m_H},
\end{equation}
where $k$ is the Boltzmann constant, $m_H$ is the mass of the
hydrogen atom,
and $y$ is the ratio of atomic to molecular hydrogen,
given by
\begin{equation}
y = \frac{\rho(H)}{\rho X},
\end{equation}
where $\rho(H)$ is the density of atomic hydrogen. In the
equilibrium state, $y$ is determined by
\begin{equation}
\frac{y^2}{1-y} =
\frac{2.11}{\rho X}\exp\left[{\frac{52490}{T}}\right]
\end{equation}
\citep{b0028}.

$u(H_2)$ is the specific internal energy of molecular hydrogen,
given by
\begin{equation}
\label{uh2start}
u(H_2) = \frac{X(1-y)E(H_2)}{2m_H}.
\end{equation}
Here $E(H_2)$ is
the energy of hydrogen molecules, and is composed of three terms,
\begin{equation}
\label{eh2}
E(H_2) = \frac{3}{2}kT + E_{rot} + E_{vib}.
\end{equation}
The terms of equation (\ref{eh2}) are the translational, rotational
and vibrational energies, respectively. The rotational energy of
hydrogen molecules is composed of the contributions of
$ortho-$ and $para-$H$_2$, and is given by
\begin{equation}
\label{urot}
E_{rot} = k\frac{(1-f_o)z_pT^2\frac{\partial\ln z_p}{\partial T}
+ f_oz_oT^2\frac{\partial\ln z_o}{\partial T}}{(1-f_o)z_p+f_oz_o},
\end{equation}
where $f_o$ is the fraction for $ortho-$H$_2$, and is $3/4$ in the
equilibrium state. It is very hard to know the exact ratio between
$ortho-$ and $para-$H$_2$ in the star forming core,
so we have used the equilibrium value in our simulations.
$z_o$ and $z_p$ in equation (\ref{urot}) are the partition
functions for $ortho-$ and $para-$H$_2$, respectively, and are given by
\begin{eqnarray}
z_p &=& \sum^\infty_{j=0,2,4,\cdots}
(2j+1)\exp\left[\frac{-j(j+1)\theta_{rot}}{T}\right],\\
z_o &=& \sum^\infty_{j=1,3,5,\cdots}
(2j+1)\exp\left[\frac{-j(j+1)\theta_{rot}}{T}\right],
\end{eqnarray}
where $\theta_{rot}$ is $85.4$K.
The vibrational energy for hydrogen molecules is given by
\begin{equation}
\label{uh2end}
E_{vib} = \frac{k\theta_{vib}}{\exp(\theta_{vib}/T)-1},
\end{equation}
where $\theta_{vib}$ is $6100$K.

The third term of equation (\ref{utotal}) is the dissociation
energy of hydrogen molecules, and is
\begin{equation}
u(H_{2diss}) = \frac{1}{2}\frac{XyD}{m_H},
\end{equation}
where $D$ is the energy of dissociation for one molecule,
and is $4.4773$eV.

$u(He)$ and $u(M)$ are the energies of helium and metals,
respectively, and are given by 
\begin{equation}
u(He) = \frac{3}{2}Y\frac{kT}{4m_H},
\end{equation}
\begin{equation}
\label{eneum}
u(M) = \frac{3}{2}Z\frac{kT}{A_m m_H},
\end{equation}
where $A_m$ is the mean atomic number of metals, and is set to 
$16.78$ in our simulation \citep{b0027}.

The specific internal energy of the gas is determined uniquely
with equations (\ref{utotal}) - (\ref{eneum}) at a given
temperature, and vice versa. However, an extra iteration is
needed to derive the temperature from a given specific internal
energy. Therefore, we have used $T$ as an independent variable
rather than $u$ in our code.

\section{Test for a nonisothermal collapse}
\subsection{Initial conditions}
MB93 developed an FDM code for radiation hydrodynamics,
and performed simulations for a centrally condensed cloud. 
They used Cartesian and spherical codes, and the Eddington
approximation for the treatment of radiative transfer.
We have performed the same test
of MB93 to compare our results. The comparison should be
meaningful because the method of MB93 and ours are
based on two completely different philosophies but deal with
exactly the same problem.

The same initial conditions have been
used in order to compare the results directly. The
initial cloud has a mass of
$1.087$M$_\odot$ and a radius of $1.1\times 10^{16}$cm.
Solidbody rotation has been imposed around the $z$--axis,
and the angular velocity is $8.2\times 10^{-12}$s$^{-1}$.

The cloud is initially spherical but centrally condensed,
and its density profile is given by
\begin{equation}
\rho = \frac{\rho_i}{\sqrt{x^2+4y^2+4z^2}},
\end{equation}
where $\rho_i = 4.28\times 10^{-16}$g/cm$^3$.
To implement this density profile, we have changed the mass of
the particles according to their position. The mass of an individual
particle is given by
\begin{equation}
m = \frac{m_i}{\sqrt{x^2+4y^2+4z^2}},
\end{equation}
where $m_i = 1.087$M$_\odot/{\cal N}_{total}$, and ${\cal N}_{total}$
is the total number of particles used in the simulation. We have used
$50000$ particles. If the isothermal collapse stage lasts up to
$\rho \sim 10^{-13}$g/cm$^3$, this is a sufficient number of
particles\footnote{We assume that all particles have the
same mass for checking the numerical Jeans condition.} 
to satisfy the numerical Jeans condition \citep{0289,0311,0161,0306,0312}.
MB93 chose this initial density
profile to see the effect of radiative transfer and heating
immediately, and to mimic the prolateness in star forming
cores \citep{0046}.

\subsection{Equation of state}
The temperature of the cloud is set to $10$K initially, and the thermal
evolution during the collapse has been tracked using the
implicit radiative transfer method explained in section
\ref{implicitsec}. For comparison, we have performed three tests.
The only
difference between each test is the calculation of the specific
internal energy. Tests 1 and 2 use the energy
equation of state of an ideal gas
for the derivation of the energy, which is given by
\begin{equation}
u = \frac{1}{\gamma-1} \frac{kT}{\mu m_h},
\end{equation}
where $\mu$ is the mean molecular weight, and is given by
\begin{equation}
\mu = \left[\frac{X(1-y)}{2}+\frac{Y}{4}+\frac{Z}{16.78}\right]^{-1}.
\end{equation}
We have used $\gamma = 5/3$ and $7/5$ in Tests 1 and 2, respectively.
The composition of the cloud is $X=0.70, Y=0.27$, $Z=0.03$ and
$y=0$,
so $\mu(=2.385)$ is assumed to be constant. In Test 3, the more realistic
energy calculation explained in section \ref{thermosec} has been used.
Note that $\mu$ is not constant in Test 3 due to the variation
of $y$.

In all tests, the gas pressure, $P$ is derived from
the pressure equation of state,
\begin{equation}
P = \frac{\rho kT}{\mu m_H}.
\end{equation}

\subsection{Results}
\begin{figure}[htbp]
\centering
\includegraphics[scale=0.4]{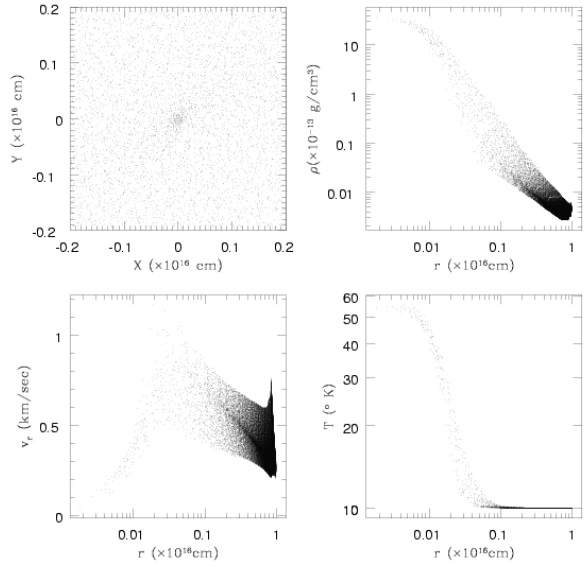}
\caption{Results for TEST 1 $(\gamma=5/3)$
at $t=0.5014t_{ff}$. The top--left
plate shows the particle positions near the center of the cloud.
An elongated central core can be seen. The density
profile (top right) of the core is nearly flat, and there is an
accretion shock around the core in the velocity plot (bottom left). The
outer envelope of the cloud remains isothermal
$(\simeq 10$K$)$ in the temperature plot (bottom right).\label{0.5014}}
\end{figure}

\begin{figure}[htbp]
\centering
\includegraphics[scale=0.4]{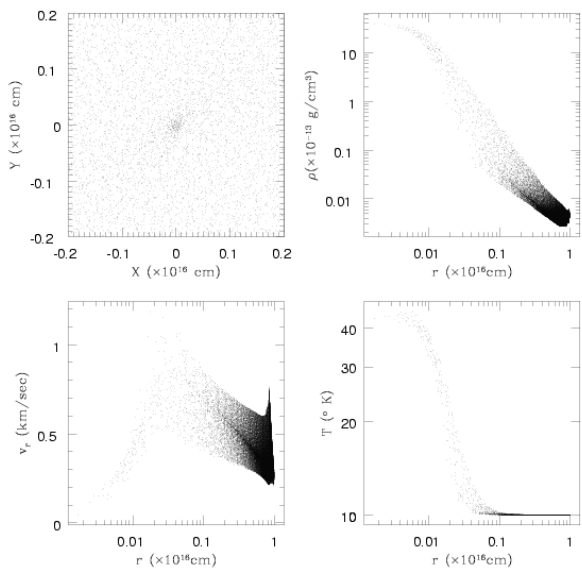}
\caption{The results for TEST 2 $(\gamma=7/5)$
at $t=0.5014t_{ff}$. The overall
features are very similar to those of Test 1, but the
temperature of the central core is lower because of the
smaller $\gamma$ value.\label{0.5014_2}}
\end{figure}

\begin{figure}[htbp]
\centering
\includegraphics[scale=0.4]{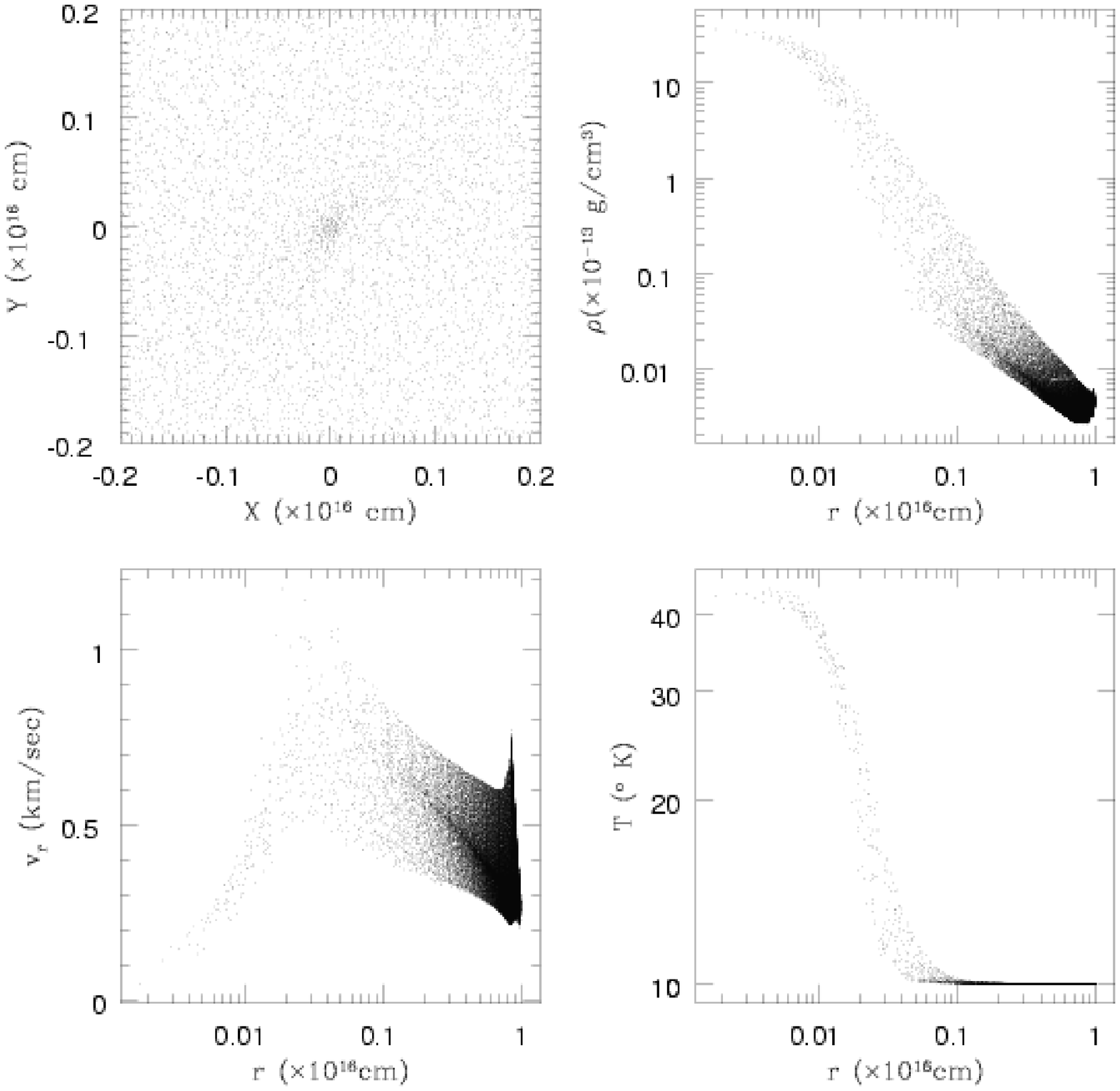}
\caption{The results for TEST 3 (variable $\gamma)$
at $t=0.5016t_{ff}$. They are more
similar to those of Test 2.\label{0.5016}}
\end{figure}

\begin{figure}[htbp]
\centering
\includegraphics[scale=0.4]{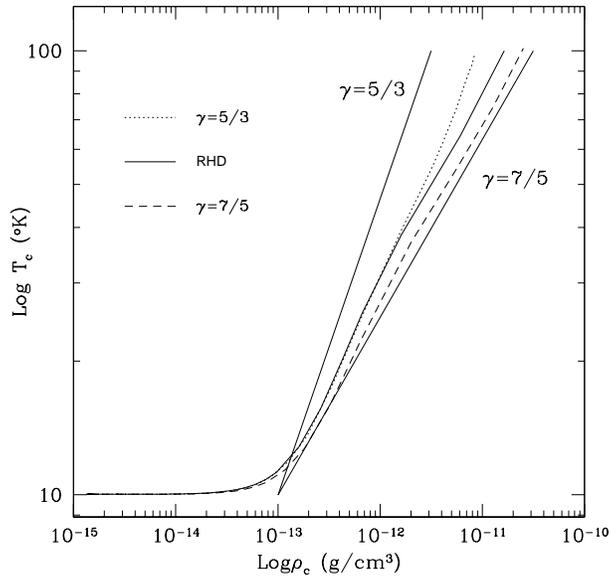}
\caption{Evolution of the density and temperature of the cloud
center. Dotted, long--dashed and solid lines are the results of
Tests 1, 2 and 3, respectively. The temperature of the collapsing
cloud is $T \simeq 10$K until $\rho_c
\simeq 10^{-13}$g/cm$^3$, and then increases afterwards.
The effective $\gamma$ value for Test 3 is nearly $5/3$ in 
$10^{-13}$g/cm$^3<\rho_c<10^{-12}$g/cm$^3$, and then changes to
$7/5$, due to the excitation of the rotational energy of 
hydrogen molecules. All tests are stopped
when the first core starts to expand.\label{rhot}}
\end{figure}

Figures \ref{0.5014} -- \ref{0.5016} show the results of Tests 1,
2 and 3, respectively. They are snapshots at
$t\simeq 0.501t_{ff}$\footnote{The freefall time
$(\simeq 3.38 \times 10^3 yrs)$ is evaluated with the assumption of
a uniform density.}. In the early stages of the collapse,
the cloud contracts
isothermally, and an elongated core forms in the center
immediately due to the initial central condensation. The central
core starts to trap the radiation and becomes adiabatic.
The transition density from the isothermal to the
adiabatic regimes is $\sim 10^{-13}$g/cm$^3$. This transition density is
nearly the same in all simulations (See Figure \ref{rhot}).
In the adiabatic stage, the temperature of the central core
tends to increase quickly,
while the outer parts of the collapsing cloud is still isothermal.
The adiabatic core is easily distinguished in the
temperature profile in Figures \ref{0.5014} -- \ref{0.5016}. The
density is flatter in the core than in the outer parts of the cloud,
and the infalling velocity drops suddenly at the edge of the
core to form a shock wave.
There is no big difference between the results of Tests 1, 2 and
3. However, the temperature of the central core in Test 1 is higher
than that of Tests 2 and 3. 
The increase of the central temperature in Test 1 is faster than
in the other simulations because of the higher $\gamma$ value.

\begin{figure}[htbp]
\centering
\includegraphics[scale=0.4]{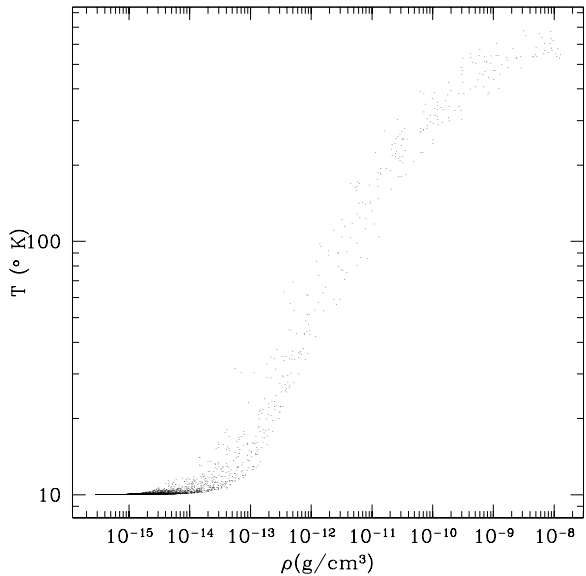}
\caption{The $\rho-T$ relation for all particles
in Test 3 at $t=0.5309t_{ff}$. The overall trend coincides well
with that of Figure \ref{rhot}, but all particles are not on a
line. The dispersion in the particle distribution 
is due to the thermal interaction between the hot central
core and its surroundings, so some particles in the density
range $10^{-14}$g/cm$^3 \sim 10^{-13}$g/cm$^3$ show a higher
temperature than the boundary value $(=10$K$)$.\label{rt}}
\end{figure}

Figure \ref{rhot} shows the evolution of the central core in
the $\rho-T$ plane. For
comparison, we have drawn two straight lines which show the slopes for 
$\gamma=5/3$ and $7/5$. When the central core
enters the adiabatic regime, Tests 1 (dots) and 2 (long--dash) 
show slightly
different evolutions. The slope of Test 1 is steeper than that of
Test 2, so the temperature in Test 1 should be higher
for the same density.
In Test 3 (solid line), the evolution of central core
follows $\gamma=5/3$ initially, but it becomes closer to
$\gamma=7/5$ when the density becomes greater than $\sim
10^{-12}$g/cm$^3$. This change in slope is due to changes in the energy
of hydrogen molecules. The rotational energy
of hydrogen molecules is not important at very low 
temperatures, so the contribution of the translational energy is
dominant. However, as the temperature increases,
the rotational energy becomes more important,
so the $\gamma$ value gets closer to $7/5$.
Could one use the $\rho-T$ relation for the cloud
center in Figure \ref{rhot} and apply it throughout the cloud,
or even to other calculations?
Figure \ref{rt} shows the $\rho-T$ relation for all particles
in Test 3 at $t=0.5309t_{ff}$. 
The temperature of some particles in the density range of
$10^{-14}$g/cm$^3 \sim 10^{-13}$g/cm$^3$ is higher than the
boundary value ($= 10$K). Furthermore,
there is a temperature dispersion at a given density through
the cloud except at the lower densities. It is due
to the non--spherically symmetric collapse, so particles closer
to the central core are hotter. If a barotropic equation of
state \citep{0452,0505} were used in the simulation, all particles
would lie on a line in the $\rho-T$ plane without any dispersion,
because there would be no thermal interaction
between the hot central core and its surroundings.
\subsection{Comparison with MB93}
We compare our results with those of MB93
in Table \ref{compmb}.
CC and SC in the table mean Cartesian and Spherical codes,
respectively. There are small differences in the results,
the temperatures of SPH are lower than those of MB93.
We may presume some reasons for this discrepancy. First of all,
the treatment for radiative transfer is different.
We have used the diffusion approximation
in the simulations, while MB93 used the Eddington
approximation. It is not easy to predict the resultant
difference due to the different treatments for the radiative
transfer,
but the diffusion
approximation may reduce the temperature increase
\citep{0265}.

Secondly, there is a difference in the energy calculation of
hydrogen molecules. We have used equations (\ref{uh2start}) -
(\ref{uh2end}), but MB93 used slightly different forms
for $u(H_2)$.
According to the $u(H_2)$ calculation of \citet{0514}
(See Appendix B of \citet{0514}), the transition temperature
from $\gamma=5/3$ to $7/5$ is $100$K. However, in our calculation
the transition temperature is variable, and $\simeq 40$K in
Test 3. Therefore, the temperature increase should be slower
in our simulation.
\begin{table}
\caption{The central density and temperature}
\label{compmb}
\begin{center}
\begin{tabular}{lcc}
\hline
case & $\rho_c=2.2\times 10^{-12}$g/cm$^3$
     & $\rho_c=1.7\times 10^{-12}$g/cm$^3$\\
\hline
Test 1 & $49.9$K & $42.6$K \\
Test 2 & $39.2$K & $34.3$K \\
Test 3 & $41.0$K & $36.8$K \\
CC & $67.0$K & \\
SC & & $56.0$K \\
\hline
\end{tabular}
\end{center}
\medskip
Comparison of our results with those of MB93. Here 
CC and SC mean Cartesian and Spherical codes, respectively.
\end{table}
\section{Summary}

We have presented a fully three-dimensional radiation hydrodynamic code 
based on the SPH method and the diffusion approximation. The difficulty
encountered in previous attempts of how to treat the double derivative
in SPH was solved by using the treatment developed by B85
to convert the double derivative to a single one. A thermal conduction
test shows that this treatment works as expected.

The second difficulty arises from the large difference between the
radiative and dynamical time scales. The radiative time scale is much
shorter than the dynamical time scale in a collapsing cloud, especially
in the isothermal stage. We have developed a fully three-dimensional
implicit scheme for dealing with the large difference
between the time scales.

To test our implicit scheme, we performed a nonisothermal cloud collapse
of a centrally condensed cloud. The same
simulation has been performed by MB93 using two FDM codes,
and we have compared our results with those of MB93.
The two numerical methods based on two completely different
philosophies agree with each other. 

\section*{Acknowledgments}
We thank Harold Yorke for providing a table of the new extinction
values for grains and for a useful discussion. This research was
supported by a grant from the National Sciences and Engineering
Research Council of
Canada. SHC thanks the Carnegie Institution of Canada for
supporting his post-doctoral fellowship at Universit\'e de Montr\'eal.

\bibliographystyle{apj}

\end{document}